\documentstyle[prl,aps, multicol,epsf]{revtex}

\begin{document}
\draft
\preprint{{\bf ETH-TH/97-15}}

\title{Characteristics of First-Order Vortex Lattice Melting: \break
  Jumps in Entropy and Magnetization}

\author{Matthew J.\ W.\ Dodgson$^{a\,}$, Vadim B.\ 
  Geshkenbein$^{a\,,b}$, Henrik Nordborg$^{a\,,c}$, and Gianni
  Blatter$^{a\,}$}

\address{$^{a\,}$Theoretische Physik, ETH-H\"onggerberg, CH-8093
  Z\"urich, Switzerland}

\address{$^{b\,}$L. D. Landau Institute for Theoretical Physics,
  117940 Moscow, Russia}

\address{$^{c\,}$Argonne National Laboratory, 9700 South Cass 
Avenue, Argonne, Illinois 60439, USA}

\date{May 16, 1997}
\maketitle
\vspace*{-0.5truecm}

\begin{abstract}
\begin{center}
\parbox{14cm}{
  
  We derive expressions for the jumps in entropy and magnetization
  characterizing the first-order melting transition of a flux line
  lattice.  In our analysis we account for the temperature dependence
  of the Landau parameters and make use of the proper shape of the
  melting line as determined by the relative importance of
  electromagnetic and Josephson interactions. The results agree well
  with experiments on anisotropic Y$_1$Ba$_2$Cu$_3$O$_{7-\delta}$ and
  layered Bi$_2$Sr$_2$Ca$_1$Cu$_2$O$_8$ materials and reaffirm the
  validity of the London model.
}

\end{center}

\end{abstract}
\vspace*{-0.5truecm}

\begin{multicols}{2}
\narrowtext

A cornerstone of the phenomenology of type II superconductors is the
Abrikosov mean-field $H$ -- $T$ phase dia\-gram comprising a
Meissner-Ochsenfeld phase at low fields, a Shubnikov- or mixed phase
at intermediate fields, and a normal metallic phase at high fields $H
> H_{c_2}$ \cite{Tinkham}. Thermal fluctuations may modify this
mean-field picture considerably as the vortex lattice melts into a
flux-liquid phase, which is the case in high temperature
superconductors \cite{Nelson,HoughtonPelcovitsSudbo}.  According to
standard symmetry considerations \cite{Landau}, this melting
transition is expected to be of first order.  Recently, this
expectation has been confirmed through the experimental observation of
a jump in the magnetization in layered Bi$_2$Sr$_2$Ca$_1$Cu$_2$O$_8$
(BiSCCO) \cite{Pastoriza,Zeldov} and in anisotropic
Y$_1$Ba$_2$Cu$_3$O$_{7-\delta}$ (YBCO) \cite{Welp} crystals. The
latent heat released in the transition has been determined in
calorimetric measurements on an YBCO crystal \cite{Schilling,Junod}
and the thermodynamic consistency between the magnetic and
calorimetric experiments via the Clapeyron relation has been
demonstrated~\cite{Schilling}.

Though consistency could be achieved in the experiments, simple
estimates \cite{Zeldov} for the jumps in magnetization and entropy
have failed to explain the magnitude and the temperature dependence of
these characteristic quantities. In particular, in the layered BiSCCO
material the entropy jump per vortex per layer as extracted from the
magnetization data via the Clapeyron equation seems to diverge upon
approaching the superconducting transition temperature \cite{Zeldov},
and no explanation of this striking result has been given so far. In
this letter we resolve this puzzle and derive the temperature
dependence of the jumps in magnetization and entropy. Our results are
consistent with all the data measured so far in YBCO and BiSCCO single
crystals, see~Fig.~1.

Besides construction of simple estimates, the theoretical discussion
of the jumps in magnetization and entropy has concentrated on
numerical simulations. The problem of the notoriously small entropy
jumps obtained in early simulations
\cite{SasikStroud,NordborgBlatterUP} has been cured by the insight of
Hu and MacDonald \cite{HuMacDonald}, who pointed out the relevance of
the temperature dependence of the Ginzburg-Landau parameters in the
analysis of numerical data obtained in simulations based on the lowest
Landau level approximation. Their insight can be given a much wider
perspective, both in terms of its application to other numerical
approaches \cite{NordborgBlatter} as well as analytic treatment of the
problem: combining this concept with our knowledge of the shape of the
melting line, we show below how to construct a consistent scheme which
provides us with expressions for the jumps in entropy and
magnetization, exhibiting both the correct magnitude and temperature
dependence. Our analysis then removes recent doubts
\cite{HuMacDonald,Schillingani} on the ability of a simple London
model to describe the large entropy jumps measured on YBCO and BiSCCO
crystals close to $T_c$.

Two important elements in our analysis are the shape of the melting
line $H_{\rm m}(T)$ and the relevant fluctuation mode in the vortex
system at the melting transition, the latter defining the volume
$V_{\rm edf}$ of the elementary degree of freedom.  Both quantities
depend sensitively on the degree of anisotropy/layeredness of the
material through the relative importance of the two types of
interactions appearing in the vortex system, the Josephson- and the
electromagnetic interaction \cite{BlatterGeshLarNord}.  As a
consequence, the melting process of the vortex lattice in YBCO and
BiSCCO exhibits quite distinct characteristics, both regarding the
shape of the melting line and the temperature dependence of the jumps
in magnetization and entropy. In addition, we will make use of the
Clausius-Clapeyron equation relating the jumps in the entropy density
$\Delta s$ and in the magnetic induction $\Delta B$,
\begin{equation}
  \Delta s = -\frac{1}{4\pi} \frac{d H_{\rm m}}{dT} \Delta B.
\label{CC}
\end{equation}

In the following, we first present the analysis for an anisotropic
material such as YBCO, where we make use of the scaling form of the
London free energy functional which contains the flux lattice constant 
as the only length scale, rendering the calculation essentially exact.
For strongly layered materials (e.g., BiSCCO), two new length scales,
the layer separation $d$ and the London penetration depth $\lambda$,
become relevant.  We show how to construct estimates for the jumps
which reduce to the previous results in the scaling regime and which
allow us to deal with layered materials.

Consider the statistical mechanics of a system described by an
effective free energy functional ${\cal F} (T, \phi)$. 
The coarse grained effective field $\phi$ (e.g.\ a Ginzburg--Landau 
wavefunction) is
obtained after integration over the microscopic degrees of freedom,
which generates the
temperature dependence in ${\cal F}$ (e.g. the Landau parameter  
$\alpha(T) = \alpha (1-T/T_c)$).
After integration over the remaining degrees of freedom $\phi$ we
arrive at the partition function ($k_{\scriptscriptstyle B}$ is the
Boltzmann constant)
\begin{equation}
  Z = \int {\cal D} \phi \, e^{-{\cal F} (T, \phi) /
    k_{\scriptscriptstyle B} T}.
\label{partition}
\end{equation} 
Taking the derivative of the free energy $F=-k_{\scriptscriptstyle B}
T \ln Z$ with respect to $T$ we obtain the entropy,
\begin{equation}
  S = S_\circ - \langle \partial_T {\cal F}\rangle,
\label{entropy}
\end{equation} 
where $S_\circ = (\langle {\cal F}\rangle - F)/T$ is the
configurational entropy of the coarse grained $\phi$-field, whereas
the second term accounts for the internal temperature dependence in
the free energy functional ${\cal F}(T, \phi)$. To be specific, let us
consider a vortex system within the London approximation where the
free energy functional takes the form
\begin{equation}
  {\cal F}[\{{\bf s}_\mu\}] = \varepsilon \varepsilon_\circ 
  a_\circ \sum_{\mu,\nu} \int d{\bf s}_\mu \cdot d{\bf s}_\nu 
  \frac{e^{-(a_\circ / \lambda) | {\bf s}_\mu-{\bf s}_\nu |}} 
  {|{\bf s}_\mu-{\bf s}_\nu|},
\label{London}
\end{equation}
with $\varepsilon_\circ = (\Phi_\circ/4\pi\lambda)^2$ the basic energy
scale in the problem (proportional to the vortex line energy),
$\lambda$ the London penetration depth, $\varepsilon^2 = m/M < 1$ is
the anisotropy parameter, and $\Phi_\circ = hc/2e$ denotes the flux
quantum. In (\ref{London}) all lengths are measured in units of the
vortex separation $a_\circ = \sqrt{\Phi_\circ/B}$, with the
configurations expressed through the dimensionless position variables
${\bf s}_\mu$. For the vortex system, the parameters
$\varepsilon_\circ$ and $\lambda$ depend on the temperature via
$\lambda^2(T) = \lambda_0^2 /[1-(T/T_c)^2]$. In the limit $\lambda >
a_\circ$ the functional (\ref{London}) assumes a simple scaling form,
with all physical parameters appearing in the prefactor $\varepsilon
\varepsilon_\circ(T) a_\circ$, the remaining factor representing a
scale independent summation over geometrical configurations of lines
\cite{review}. Within the scaling regime, all physical results depend
on the combination $\varepsilon \varepsilon_\circ(T) a_\circ$. E.g.,
the shape $B_{\rm m} (T)$ of the melting line is given by the
condition $\varepsilon \varepsilon_\circ(T_{\rm m}) a_\circ/
k_{\scriptscriptstyle B} T_{\rm m} = {\rm const.}$ ($= 1/2 \sqrt{\pi}
c_{{\rm\scriptscriptstyle L}}^2$, we make use of the usual definition
of the Lindemann number $c_{\scriptscriptstyle L}$), from which one
easily derives the standard result
\cite{HoughtonPelcovitsSudbo,BlatterGeshLarNord}
\begin{equation}
  B_{\rm m} (T) \approx \frac{\Phi_\circ}{\lambda^2} 4\pi
  c_{{\rm\scriptscriptstyle L}}^4 \frac{\varepsilon^2
    \varepsilon_\circ^2 \lambda^2}{(k_{\scriptscriptstyle B}T)^2}
  \propto \biggl(1-\frac{T^2}{T_c^2}\biggr)^2
\label{bj}
\end{equation}
(note that Eq.~(\ref{bj}) derives from scaling and proves the validity
of the Lindemann criterion). Returning to (\ref{entropy}), we can
again make use of the scaling form of ${\cal F}$ and express the
second term in (\ref{entropy}) through the energy $\langle {\cal
  F}\rangle$: $\langle \partial_T {\cal F}\rangle = (\partial_T \ln
\varepsilon_\circ) \langle {\cal F}\rangle$ and inserting back into
(\ref{entropy}) we arrive at the relation
\begin{equation}
S = S_\circ \left( 1-\frac{T}{\varepsilon_\circ}
\frac{d \varepsilon_\circ}{dT} \right) + \frac{T}{\varepsilon_\circ}
\frac{d \varepsilon_\circ}{dT} \frac{F}{T}.
\label{entropy2}
\end{equation}
At the phase transition the entropy exhibits a jump while the free
energy $F$ remains continuous. From (\ref{entropy2}) we thus infer the
following relation between the entropy jump $\Delta S$ and its
configurational part $\Delta S_\circ$,
\begin{equation}
  \frac{\Delta S}{\Delta S_\circ} = \left(1-\frac{T}
  {\varepsilon_\circ} \frac{d \varepsilon_\circ}{dT} \right) 
  = \frac{1+(T_{\rm m}/T_c)^2}{1-(T_{\rm m}/T_c)^2},
\label{entropyjs}
\end{equation}
where we have made use of the specific temperature dependence of the
line energy $\varepsilon_\circ \propto [1-(T/T_c)^2]$ in the last
equation. We find, that close to the thermodynamic transition the
entropy jump $\Delta S$ is strongly enhanced with respect to its
configurational component $\Delta S_\circ$. Note that it is the latter
quantity which is amenable to simple estimates \cite{Zeldov} and which
is usually calculated in numerical simulations \cite{NordborgBlatter}.
The result (\ref{entropyjs}) can be easily understood as the simple
consequence of a ``coordinate" transformation: the temperature enters
the partition function $Z$ not merely as a scale parameter but rather
in a combination $T/[1 - (T/T_c)^2]$. Close to $T_c$ this expression
becomes singular, resulting in a strong enhancement of the entropy
jump \cite{HuMacDonaldcom}. The physical origins of this enhanced
entropy are microscopic
fluctuations: Within the
coarse grained vortex model these fluctuations surface in the
temperature dependence of the phenomenological parameters in
${\cal F}$, which are singular at the mean--field transition.

Next, let us find an expression for the configurational part $\Delta
S_\circ$ of the entropy jump: within the scaling regime, standard
arguments {\it dictate} the form
\begin{equation}
  \Delta S_\circ = \eta k_{\scriptscriptstyle B} \frac{V}{V_{\rm edf}},
\label{centropyj}
\end{equation}
with $V_{\rm edf} = \varepsilon a_\circ^3$, $V$ denotes the system
volume, and $\eta$ is a small number. Physically, this result can be
understood by attributing the thermal energy $k_{\scriptscriptstyle B}
T$ to each individual degree of freedom. The volume $V_{\rm edf}$ is
determined by the dominant modes leading to melting, which are located
at the Brillouin zone boundary with $k_\perp \approx \sqrt{4
  \pi}/a_\circ$. Hence, we can define the volume per degree of freedom
in the form $V_{\rm edf} = a_\circ^2 L$. Furthermore, in an
anisotropic material the important fluctuations involve the wavevector
$k_z \sim 1/\varepsilon a_\circ$ along the field, thus $L =
\varepsilon a_\circ$ (with all numericals absorbed in $\eta$).
Combining (\ref{entropyjs}) and (\ref{centropyj}) we obtain the final
result for the entropy jump per vortex per layer,
\begin{equation}
\Delta S_d \approx 2 \eta \frac{d}{\varepsilon a_\circ}
  \frac{k_{\scriptscriptstyle B}}{1-(T_{\rm m}/T_c)^2}.
\label{dsj}
\end{equation}
On the melting line, the product $a_\circ [1-(T_{\rm m}/T_c)^2]$ is
(roughly) temperature independent and we find a {\it constant} but
material dependent entropy jump per vortex per layer, in agreement
with experiments on YBCO \cite{Schilling}.

In order to find a numerical result for the entropy jump we compare
the latent heat per vortex line $L_l = T_{\rm m} \Delta S_\circ
a_\circ^2$ with the result of numerical simulations carried out within
the present London formalism \cite{NordborgBlatter}: Making use of
(\ref{bj}) and the numerical result $L_l=0.015\,\varepsilon_\circ$ we
find the expression $\Delta S_d \approx 0.03 \, \varepsilon_\circ(0)
d/T_c$ and using parameters for YBCO ($\lambda \approx 1400 $ \AA~ and
$d = 12$ \AA) we obtain the value $\Delta S_d \approx 0.4\,
k_{\scriptscriptstyle B}$, in good agreement with experiment
\cite{Schilling}. Using the result for the melting line in
Ref.~\cite{NordborgBlatter}, $\varepsilon \varepsilon_\circ(T_{\rm m})
a_\circ/ k_{\scriptscriptstyle B} T_{\rm m} \approx 11$, we obtain a
value for the parameter $\eta$, $\eta \approx 0.16$.

Let us turn next to the jump $\Delta B$ in the induction. In order to
make use of the Clapeyron equation (\ref{CC}) we have to determine the
slope $\partial_T B_{\rm m}$ of the melting line \cite{ignore}.
Ignoring for the moment the temperature dependence in the free energy
(\ref{London}) (and thus in the parameters of (\ref{bj})) we have
$\partial_T B_{\rm m} = -2 B_{\rm m} / T_{\rm m}$. The full result
which accounts for the temperature dependence in $\varepsilon_\circ$
reads
\begin{equation}
\frac{d B_{\rm m}}{d T} = - \frac{2 B_{\rm m}}{T_{\rm m}}
\left(1-\frac{T}{\varepsilon_\circ}
  \frac{d \varepsilon_\circ}{dT} \right).
\label{slopes}
\end{equation}
The same factor relating the two slopes appears in equation
(\ref{entropyjs}) which expresses the entropy jump $\Delta S$ through
$\Delta S_\circ$. Thus in the final expression for the jump $\Delta B$
in the induction this correction factor drops out and we arrive at the
simple result (we use $V_{\rm edf} = a_\circ^2 L$)
\begin{equation}
  \Delta B = \mu \frac{k_{\scriptscriptstyle B} T_{\rm m}}
  {\Phi_\circ L},
\label{dbgen}
\end{equation}
where $\mu = 2 \pi \eta \approx 1.0$. Using $L = \varepsilon a_\circ$
as appropriate in an anisotropic material we arrive at the final
result for the jump in $B$,
\begin{equation}
  \Delta B \approx \mu \frac{k_{\scriptscriptstyle B} T_{\rm m}}
  {\Phi_\circ \varepsilon a_\circ} \approx 6. \, 10^{-4} \,
  \frac{\Phi_\circ}{\lambda^2(T_{\rm m})}.
\label{dbj}
\end{equation}
Note that in an incompressible (uncharged, $e \rightarrow 0$ and
$\lambda \rightarrow \infty$) system, we correctly find $\Delta B
\rightarrow 0$. Rewriting (\ref{dbj}) in the form $\Delta B [{\rm G}]
\approx (1.5 \cdot 10^{-6}/\varepsilon) T_{\rm m}[{\rm K}] (B_{\rm
  m}[{\rm G}])^{1/2}$ and choosing $\varepsilon = 1/8$ we arrive at a
good agreement with the magnetization data of Schilling {\it et al.}
\cite{Schilling} on an YBCO single crystal, see Fig.~1 (we have made
use of the experimentally measured melting line $B_{\rm m}(T)$).

So far our analysis has been essentially exact: we have exploited the
scaling behavior of the London functional in the regime $a_\circ <
\lambda$ and have deduced the single unknown parameter $\eta$ from a
comparison to a numerical simulation of the London model. While this
approach is successfully applied to a continuous anisotropic
superconductor such as YBCO, we have to reconsider the situation for
strongly layered materials (e.g., BiSCCO) as new length scales ($d$,
$\lambda$) enter the problem. We then can make use of an important
insight provided by the above derivation, namely that the calculation
of the jump in $B$ does not suffer from the complications associated
with the determination of the jump in entropy. This is because the
entropy involves a derivative of the free energy with respect to
temperature, whereas the induction is given by the derivative with
respect to the magnetic field.  The latter usually does not show up in
the Landau parameters \cite{spoils}. Indeed, we can arrive at the
result (\ref{dbgen}) starting from the general thermodynamic relation
$B = - (4\pi/V) \partial_H G|_T$, where $G$ is the Legendre transform
of the free energy $F$, $G(T,H) = F(T,B) - BHV/4\pi$. With the
estimate $G \sim k_{\scriptscriptstyle B} T V/V_{\rm edf}$ and making
use of the power law dependence of $V_{\rm edf}$ on $H$ we obtain
\begin{eqnarray}
  \Delta B &\approx& \frac{\mu' k_{\scriptscriptstyle B} T_{\rm m}}
  {H V_{\rm edf}}.
\label{Bsing}
\end{eqnarray}
Inserting the ansatz $V_{\rm edf} = a_\circ^2 L = \Phi_\circ L/H$ we 
immediately recover the result (\ref{dbgen}) 
(with $\mu$ replaced by an unknown numerical $\mu'$). 
Once the jump in the induction $\Delta B$ is known,
we can make use of the Clausius-Clapeyron equation (\ref{CC}) and
arrive at the result for the jump in the entropy. The problematic
factor arising from the temperature derivative of the free energy
functional ${\cal F}$, see (\ref{entropy}) and (\ref{entropyjs}), is
taken care of by the derivative $\partial_T B_{\rm m}$ of the melting
line, see (\ref{slopes}).

We proceed with the analysis of the jumps in magnetization and entropy
for strongly layered superconductors, following the above line of
thought. To do so we need to know the length $L$ of the relevant modes
at melting, see (\ref{dbgen}), as well as the shape of the melting
curve in a layered superconductor, to be used in (\ref{CC}).

In layered BiSCCO the melting line is pushed down to low fields
$B_{\rm m} (T) < B_\lambda (T) = \Phi_\circ / \lambda^2$ over a large
portion of the phase diagram. The dominant interaction in the vortex
system is then given by the electromagnetic one.  The loosely bound
pancake vortices undergo large thermal fluctuations and dominate the
melting process, hence $L = d$. The shape of the melting line follows
most easily from a Lindemann analysis with $\langle u^2 \rangle =
c_{\rm\scriptscriptstyle L}^2 a_\circ^2 \sim T d/\varepsilon_l (k_z
\sim 1/d)$, while making use of the dispersive electromagnetic line
tension $\varepsilon_l \sim \varepsilon_\circ/\lambda^2 k_z^2$
\cite{BlatterGeshLarNord},
\begin{equation}
  B_{{\rm m}}^{{\rm em}} (T) \approx \frac{\Phi_\circ}{\lambda^2}
  \frac{c_{{\rm\scriptscriptstyle L}}^2}{2} \frac{ \varepsilon_\circ
    d}{k_{\scriptscriptstyle B}T} \propto
  \biggl(1-\frac{T^2}{T_c^2}\biggr)^2.
\label{bem}
\end{equation}
Close to $T_c$ the Josephson interaction becomes relevant as soon as
$\varepsilon \lambda(T) > d$: For $T > T^{\rm em} \approx T_c [1 -
(\varepsilon \lambda_0/d)^2]^{1/2}$ the dominant fluctuations at
melting are cut off on the larger scale $L \sim \varepsilon \lambda$.
The Lindemann criterion takes the form $\langle u^2 \rangle =
c_{\rm\scriptscriptstyle L}^2 a_\circ^2 \sim T \varepsilon \lambda/
\varepsilon_l (k_z \sim 1/\varepsilon \lambda)$ and we obtain a
melting line following a $(1-T^2/T_c^2)^{3/2}$ behavior
\cite{BlatterGeshLarNord},
\begin{equation}
  B_{{\rm m}}^{{\rm em,J}} (T) \approx \frac{\Phi_\circ}{\lambda^2}
  \frac{\pi c_{{\rm\scriptscriptstyle L}}^2}{4} \frac{\varepsilon
    \varepsilon_\circ \lambda}{k_{\scriptscriptstyle B}T} \propto
  \biggl(1-\frac{T^2}{T_c^2}\biggr)^{3/2}.
\label{bemj}
\end{equation}
Inserting the above results for $L$ into (\ref{dbgen}) we obtain the
jump $\Delta B$ in the induction for a {\it layered} material,
\begin{equation}
  \Delta B \approx \left\{ \begin{array}{r@{\quad\quad}l}
      \displaystyle{\mu' \frac{k_{\scriptscriptstyle B} T_{\rm m}}
        {\Phi_\circ d} }, \!\! & T_{\rm m} < T^{\rm em},\\ \noalign{\vskip
        5 pt} \displaystyle{\mu' \frac{k_{\scriptscriptstyle B} 
        T_{\rm m}} {\Phi_\circ \varepsilon \lambda_0}
        \sqrt{1-(T_{\rm m}/T_c)^2} }, \!\! & T^{\rm em} < T_{\rm m}.
        \end{array}\right.
\label{dbemj}
\end{equation}
Note that it is the temperature dependence of $L$, which goes from
$\varepsilon a_\circ (T_{\rm m})$ in a continuous anisotropic
superconductor to $L = d$ and $L = \varepsilon \lambda(T_{\rm m})$ in
a layered material, that leads to the different dependencies in the
jump $\Delta B(T)$.

The result (\ref{dbemj}) explains the experimental observations of
Zeldov {\it et al.} \cite{Zeldov}, see Fig.~1: At low temperatures
$T_{\rm m} < T_c - 7$ K the jump $\Delta B$ increases linearly with
temperature. About 7 K before reaching the transition, $\Delta B$
drops sharply and vanishes at $T_c$. This behavior is explained in
terms of the crossover at $T^{\rm em}$, where the Josephson coupling
between the layers becomes relevant and cuts off the further growth of
fluctuations.  With the same value $\mu' = \mu $ ($\approx 1$)
as above, we rewrite (\ref{dbemj}) in the form $\Delta B [{\rm G}]
\approx (0.07/d[{\rm \AA}]) T_{\rm m}[{\rm K}]$ and $\Delta B [{\rm
  G}] \approx (0.07/\varepsilon \lambda_0[{\rm \AA}]) T_{\rm m}[{\rm
  K}] [1-(T_{\rm m}/T_c)^2]^{1/2}$ and using the parameters 
$d = 15~{\rm \AA}$, $\lambda_0 \approx 2000~{\rm \AA}$, 
and $\varepsilon \approx 1/400$ we obtain
excellent agreement with the experimental result of Zeldov {\it et
  al.}  \cite{Zeldov}. Note that the experimental finding $\Delta B (T
\rightarrow T_c) \rightarrow 0$ (while $B$ remains $\gg \Delta B$),
combined with (\ref{dbgen}), hints at a divergence $L (T \rightarrow
T_c) \rightarrow \infty$ and thus a 3D transition as $T_c$ is
approached. Finally, the expressions for the jumps in entropy follow
from the Clapeyron equation (using the line shapes (\ref{bem}) and
(\ref{bemj})),
\begin{equation}
  \Delta S_d \approx \left\{ \begin{array}{r@{\quad\quad}l}
      \displaystyle{\frac{\mu}{\pi} 
      \frac{k_{\scriptscriptstyle B}}{1-(T_{\rm m}/T_c)^2} }, 
      & T_{\rm m} < T^{\rm em},\\ 
      \noalign{\vskip 5 pt} \displaystyle{\frac{3 \mu}{4 \pi} 
      \frac{d}{\varepsilon\lambda_0} 
      \frac{k_{\scriptscriptstyle B}}{\sqrt{1-(T_{\rm m}/T_c)^2}} }, 
      & T^{\rm em} < T_{\rm m}. \end{array}\right.
\label{dsemj}
\end{equation}
Unlike in the anisotropic case, for the layered material the entropy
jump per vortex per layer diverges on approaching the transition at
$T_c$, again in agreement with the experimental observation
\cite{Zeldov} (note that at low temperatures the entropy jump in Ref.
\cite{Zeldov} vanishes as the melting line flattens, possibly due to
disorder).

In summary, our new estimates (\ref{dsj}), (\ref{dbj}) and
(\ref{dbemj}), (\ref{dsemj}) provide a consistent explanation of the
observed characteristic jumps at the first-order melting transition of
the vortex crystal in type II superconductors.

We thank T.~Forgan, M.~Moore, A.~Schilling, U.~Welp, and E.~Zeldov for
discussions and AS, UW, and EZ for providing us with their original
data. We gratefully acknowledge financial support from the Swiss
National Foundation. Work at Argonne was supported by the NSF-Office
of Science and Technology Centers under contract No. DMR91-20000.


\begin{figure}
  \makebox[1.75in]{\rule[1.125in]{0in}{1.125in}}
  \includegraphics{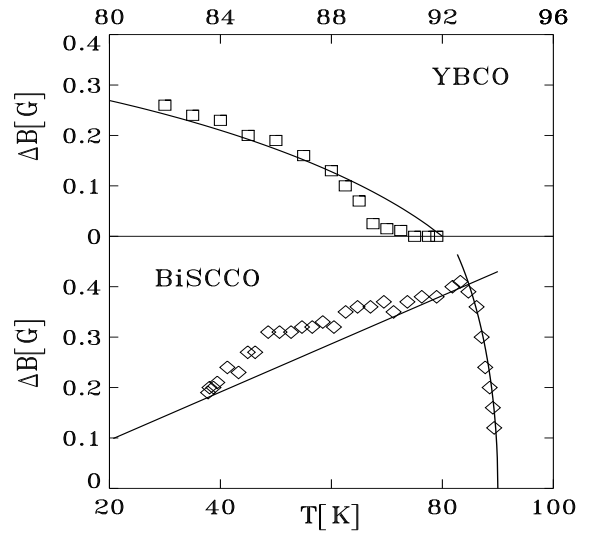} \vglue 1.8truecm

\caption{Top: the jump $\Delta B$ in the induction versus temperature $T$
  as measured in an YBCO single crystal [8] and calculated from the
  expression (12). The deviations close to the transition are possibly
  due to sample inhomogeneity [7]. Bottom: the same for a
  BiSCCO single crystal [6] using the result (16). The drop in
  $\Delta B$ on approaching $T_c$ is explained in terms of a 
  temperature dependent cutoff in the electromagnetic fluctuations 
  through the Josephson coupling at temperatures $T > T^{\rm em}$. 
  The entropy jump $\Delta S_d$ per vortex per layer as obtained 
  from $\Delta B$ through the Clapeyron equation diverges at the 
  transition [6], in agreement with the result (17).}
\label{fig:1}
\end{figure}
\end{multicols}

\begin{thebibliography}{99}
\bibitem{HuMacDonaldcom} It follows that this analysis is not
  particular to the lowest Landau level approximation as 
  claimed in Ref.~\cite{HuMacDonald}.
\bibitem{ignore} We ignore the small difference between $B$ and $H$.
\bibitem{spoils} Within the London description of type II 
superconductors the dependence on $1 - (T/T_c)^2 - H/H_{c_2}(0)$ 
modifies the melting line. However, since $H_{\rm m}/H_{c_2}$
remains small, our simplified discussion is quantitatively correct.
\end{thebibliography}
\end{document}